\documentclass[11pt]{article}
\usepackage{epsfig}
\usepackage{amsmath}
\usepackage{amssymb}

\begin{document}

\title{Can we extract the pion electromagnetic form factor from 
 a $t$-channel diagram only?}
\author{T. Mart\thanks{E-mail address: tmart@fisika.ui.ac.id}\\
{\sl Departemen Fisika, FMIPA, Universitas Indonesia, Depok, 16424, Indonesia}}

\maketitle

\begin{abstract}
We show that we are able to extract the pion 
electromagnetic form factor by using 
the recent charged pion electroproduction data from JLab and a 
simple $t$-channel diagram. For this purpose we have performed
the $Q^2$-independent and dependent analyses.
The result of the first analysis is in good agreement
with those of previous works and fit the Maris and Tandy model
as well as the monopole parameterization which describes a
pion radius of 0.672 fm. The result of the second analysis
corroborates the findings in the first analysis.
Our findings therefore provide a direct proof
that at the given kinematics the $t$-channel diagram really
dominates the process. This could also set a new constraint to the phenomenological
models that try to describe the process.\\[2ex]
\end{abstract}

\newpage

\section{\sc Introduction}
In nuclear physics pions occupy a special place. Being the
lightest meson in its family, pions play the major role in
strong interaction. In fact, pions were the first mesons
predicted by Yukawa as the mediator of the strong nuclear force.
However, despite intensive investigations for decades, our
understanding of the pion structure is far from complete.
This is reflected, e.g., by the incomplete available data base
of the pion electromagnetic form factor $F_\pi(Q^2)$ in a considerably
wide range of virtual photon momentum transfer $Q^2$. We 
understand that $F_\pi$ is nothing but a Fourier transformation 
of the charge distribution.

At very small $Q^2$ the behavior of $F_\pi$ has been accurately
determined by scattering of high energy pions from atomic 
electrons\,\cite{Amendolia:1986wj}. At higher $Q^2$ values
one should consult pion electroproduction off a nucleon.

With the continuous and excellent electron beams available
at modern accelerators, such as CEBAF at JLab, it is now
possible to measure pion electroproduction observables with
unprecedented accuracy, at certain kinematics inaccessible
with previous technology. This allows for a precise extraction
of the longitudinal differential cross section via a Rosenbluth
separation, at very small values of Mandelstam variable $-t$
\cite{volmer,Horn:2006tm,Tadevosyan:2007yd}. These recent data
show that the cross section $d\sigma/dt$ at this kinematics 
exhibit a steep decreasing function of $-t$, indicating the
dominance of the $t$-channel contribution in the process.
This behavior becomes the first pillar for the $F_\pi$
extraction in Refs.\,\cite{volmer,Horn:2006tm,Tadevosyan:2007yd}.
The extracted $F_\pi$ values within the range of 
$0.6\lesssim Q^2\lesssim 2.5$ GeV$^2$ are impressively accurate.
Nevertheless, it is important to note that these extractions:
\begin{itemize}
\item rely heavily on the $t$-channel dominant assumption, although
 the assumption has never been directly checked.
\item are model dependent. In Refs.\,\cite{volmer,Horn:2006tm,Tadevosyan:2007yd}
 the Regge model of Ref.\,\cite{vgl} has been utilized to obtain $F_\pi$.
\item are plagued with the model uncertainties, which come from the
  extrapolation of the form factor cut-off to the physical limit.
  These uncertainties have been reported by Ref.\,\cite{Tadevosyan:2007yd}.
\item are ``strategy'' dependent. As shown in Ref.\,\cite{Tadevosyan:2007yd},
by using the same data and model as in Ref.\,\cite{volmer} but focusing
the analysis on the minimal $t$ allowed by kinematics, where the
unwanted background is assumed to be negligible, 
smaller $F_\pi$ values are obtained.
\end{itemize}
The above facts indicate that current extraction methods are far
from established. Since
a completely model-independent extraction is quite difficult and
challenging (although it is not impossible, see Ref.\,\cite{deo_bisoi}
for instance), attempts in this direction would be invaluable for
helping to resolve this issue.

It is the purpose of this paper to 
revisit a simple method, \`a la Chew-Low, which is based only on 
the first assumption. This will also serve as an alternative 
method, which is obviously easy to check.

This paper is organized as follows. In Section~\ref{Formalism}
we briefly describe the formalism used in our calculation. 
Section~\ref{$Q^2$_Independent_Analysis}
presents the result of the $Q^2$-independent analysis, where
we extract the electromagnetic form factor for fixed $Q^2$ values.
Section~\ref{$Q^2$_Dependent_Analysis} demonstrates the result 
of the $Q^2$-dependent analysis, i.e., the result when we parameterize the 
form factor with a monopole form and fit all available data simultaneously.
We shall summarize and conclude 
our findings in Section~\ref{Conclusion}. 

\begin{figure}[t]
  \begin{center}
    \leavevmode
    \epsfig{figure=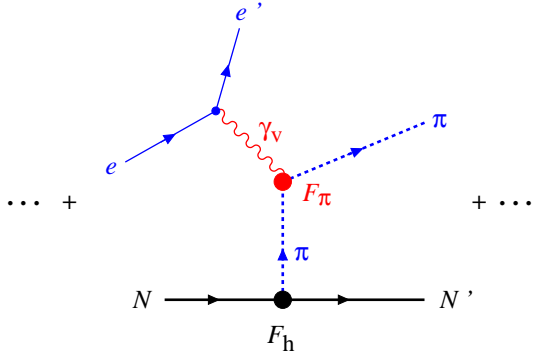,width=70mm}
    \caption{(Color online) The $t$-channel diagram that dominates the contributions
        to the pion electroproduction process. The electromagnetic
        and hadronic form factors are denoted by $F_\pi$ and $F_{\rm h}$,
        respectively.\protect\label{fig:feynman}}
  \end{center}
\end{figure}

\section{\sc Formalism}
\label{Formalism}
We begin with the convention of four-momenta given in the process
\begin{eqnarray}
\gamma_v (k) + N (p) \to \pi(q) + N(p') ~,
\end{eqnarray}
to write the $t$-channel amplitude for the diagram 
shown in Fig.~\ref{fig:feynman} as\,\cite{bjorken,fnw}
\begin{equation}
  {\cal M} = \left[\frac{2ieg_{\pi NN}F_\pi(k^2)F_{\rm h}(t)}{k^2(t-m_\pi^2)}\right]
  {\bar u}(p')\gamma_5(k^2 q\cdot\epsilon-k\cdot q k\cdot\epsilon) u(p) ,
\label{eq:a51}
\end{equation}
where $e$ is the proton charge, $\epsilon$  denotes the virtual
photon polarization, while $m_\pi$ and 
$g_{\pi NN}$ are the pion mass and the pion-nucleon coupling constant, respectively.

In Eq.~(\ref{eq:a51}) 
the hadronic and electromagnetic form factors are denoted by
$F_{\rm h}$ and $F_\pi$, respectively, while the Mandelstam variable
$t$ can be written as
\begin{eqnarray}
  \label{eq:mandelstam-t}
  t=(k-q)^2=k^2+m_\pi^2-2k\cdot q ~.
\end{eqnarray}
Note that in Eq.~(\ref{eq:a51}) we have added the so-called
the Fubini-Nambu-Wataghin term\,\cite{fnw} in order to retain 
the gauge invariance\,\cite{gauge-invariance}.

Clearly, this amplitude contributes only to the longitudinal 
cross section. For meson electroproduction it is well known that this
cross section can be written as\,\cite{deo}
\begin{equation}
\frac{d\sigma_{\rm L}}{d\Omega}=\frac{-2|\vec{q}|Wk^2}{(s-m^2)k_0^2}~
\Bigl[ |f_5|^2 + |f_6|^2 + 2 \cos \theta {\rm Re}(f_5^* f_6) \Bigr] ,
\label{eq:dcsL}
\end{equation}
where $W$ indicates the total c.m. energy, $\theta$ is the pion scattering
angle, and 
\begin{equation}
f_{5,6} = \mp \frac{k_{0}}{8 \pi W}\left(\frac{E' \pm m}{E \pm m}\right)^{1/2}
(q \cdot k k_{0} - q_0 k^{2})A_{5} ~,\label{eq:f5andf6}
\end{equation}
with $m$ denotes the nucleon mass and 
$A_5$ is given by the term inside the square bracket in the r.h.s.
of Eq.~(\ref{eq:a51}).
Inserting Eq.~(\ref{eq:f5andf6}) into Eq.~(\ref{eq:dcsL}) and by making
use of 
\begin{equation}
\frac{d\sigma_L}{dt} ~=~ \frac{\pi}{|\vec{k}|\,|\vec{q}|}\,\frac{d\sigma_L}{d\Omega}~,
\end{equation}
we obtain
\begin{equation}
\frac{d\sigma_L}{dt} = t\,\frac{e^2g_{\pi NN}^2k^2F_\pi^2(k^2)F_{\rm h}^2(t)}{
8\pi W(s-m^2)(t-m_\pi^2)^2|\vec{k}|^3}\left(q_0-\frac{k\cdot q}{k^2}k_0\right)^2.
\label{eq:dsL/dt}
\end{equation}
Since we are working with small $t$, we can neglect both
$t$ and $m_\pi$ in Eq.~(\ref{eq:mandelstam-t}) to obtain
$k^2\approx 2k\cdot q$, so that the last term of Eq.~(\ref{eq:dsL/dt}) 
can be reduced to $(q_0-\frac{1}{2}k_0)$. This has an advantage,
because we have reduced the dependencies of the cross section on $t$. Finally,
we can shift the pion propagator term to the l.h.s. of Eq.~(\ref{eq:dsL/dt}) 
and using $Q^2=-k^2$ to obtain
\begin{equation}
(t-m_\pi^2)^2\frac{d\sigma_L}{dt} = -t\,\frac{e^2g_{\pi NN}^2Q^2F_\pi^2(Q^2)
F_{\rm h}^2(t)}{8\pi W(s-m^2)|\vec{k}|^3}\,\left(q_0-{\textstyle
\frac{1}{2}}k_0\right)^2~.
\label{eq:dsL/dt*(t-m2)}
\end{equation}
The r.h.s. of Eq.~(\ref{eq:dsL/dt*(t-m2)}) were a linear function of $t$,
if we could neglect the hadronic form factor $F_{\rm h}$. If this were
the case, then an extrapolation of Eq.~(\ref{eq:dsL/dt*(t-m2)}) to the
pion pole ($t=m_\pi^2$) could be directly performed to determine the
$F_\pi$ at a certain $Q^2$ value, because the r.h.s. is free of poles and 
the extrapolation (setting $t$ at $m_\pi^2$) will eliminate other
contributions beyond the $t$-channel one.
However, as we clearly understand, this is not the case. 

To overcome the shortcoming of the extrapolation method, here
we only assume that Eq.~(\ref{eq:dsL/dt*(t-m2)}) is valid in the physical
region ($t<0$, but still small) because the $t$-channel dominates
the process. Thus, for each value of $Q^2$, Eq.~(\ref{eq:dsL/dt*(t-m2)}) can be fitted to
the data in the form of
\begin{eqnarray}
  \label{eq:fit_function}
  y=F_\pi^2\cdot f(t)+g~,
\end{eqnarray}
to obtain $F_\pi$. In Eq.~(\ref{eq:fit_function}) we introduce
the $g$ parameter to account for other terms neglected 
in Eq.~(\ref{eq:dsL/dt*(t-m2)}). Obviously, the use of this parameter 
improves the $\chi^2$ significantly, thus, absorbing more
information from the electroproduction data, but at the cost of 
increasing the flexibility of the fit, i.e., increasing the
error bars of the extracted $F_\pi$. 

\begin{figure}[!h]
  \begin{center}
    \leavevmode
    \epsfig{figure=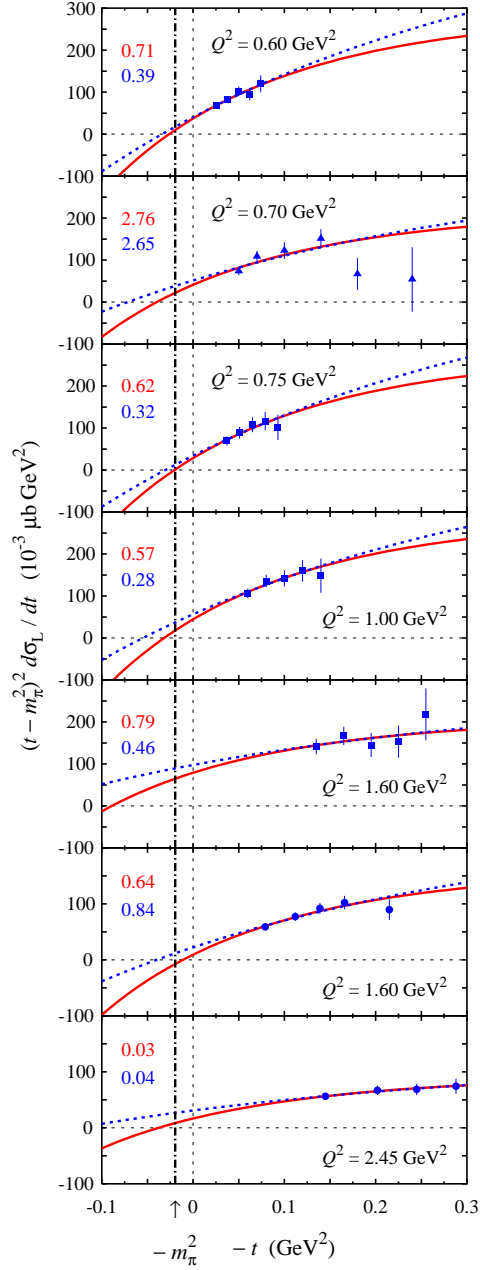,width=70mm}
    \caption{(Color online)  The modified longitudinal differential cross section
      as a function of the Mandelstam variable $-t$. Solid (dotted) lines 
      show the results of fitting to experimental data with
      $\Lambda_{\rm h}=0.85~(1.30)$ GeV. Solid squares 
      \protect\cite{Tadevosyan:2007yd} and solid circles 
      \protect\cite{Horn:2006tm} show the recent JLab data, 
      whereas solid triangles display the older
      data from DESY~\protect\cite{desy}. The upper (lower) numbers at the left part
      of each panel show the values of $\chi^2/N$ obtained for
      $\Lambda_{\rm h}=0.85~(1.30)$ GeV.
      \protect\label{fig:(t-mu2)dsig_dt}}
  \end{center}
\end{figure}

\section{\sc $Q^2$-Independent Analysis}
\label{$Q^2$_Independent_Analysis}
In this analysis we first fix the value of the pion-nucleon coupling constant to be 
$g_{\pi NN}^2/(4\pi)=14.17$\,\cite{Ericson:2000md}. We note that in the
literature this value varies from 13.45\,\cite{Timmermans} to 
14.52\,\cite{Rahm:1998jt}. Thus, our choice lies nicely in this range.
Later, we will also try to investigate the influence of the
coupling constant variation on the extracted form factor. 
For the hadronic vertex we adopt the monopole form factor
\begin{equation}
F_{\rm h}(t) = \left(1-\frac{t}{\Lambda_{\rm h}^2}\right)^{-1}~.
\end{equation}
Reference\,\cite{volmer_thesis} has pointed out that $\Lambda_{\rm h}=
0.85$ GeV is generally used, whereas the precise value of $\Lambda_{\rm h}$
is a matter of controversy and may vary between 0.4 and 1.5 GeV.
Meson exchange models for nucleon-nucleon interaction dictates that 
$\Lambda_{\rm h}\gtrsim 1.30$ GeV\,\cite{Machleid87}, whereas lattice 
QCD calculations prefer a value of 0.75 GeV\,\cite{Liu95}. On the 
other hand QCD sum rule calculations yield $\Lambda_{\rm h}=0.80$ GeV 
\cite{Mei95}, while K. Vansyoc\,\cite{vansyoc} found that 
$\Lambda_{\rm h}\lesssim 1.00$ GeV is demanded by the present pion 
electroproduction data.

In this work we found that the best result is obtained by using 
$\Lambda_{\rm h}=1.30$ GeV, i.e., the case when we consider
the result of meson exchange models for the nucleon-nucleon
interactions\,\cite{Machleid87}. Nevertheless, using the generally
accepted value (i.e. $\Lambda_{\rm h}=0.85$ GeV) the obtained result
is still in good agreement with previous analyses.

At $W=1.95$ GeV there are four sets of data points from 
Ref.\,\cite{Tadevosyan:2007yd} with $Q^2=0.6$, 0.75, 1.00
and 1.60 GeV$^2$, as a function of $t$. At higher energy 
and momentum transfers ($W=2.22$ GeV, $Q^2=1.6$ and 2.45 GeV$^2$), 
two more data sets are available from Ref.\,\cite{Horn:2006tm}. For the sake 
of comparison, we also present the reanalyzed DESY data at 
$W=2.19$ GeV, $Q^2=0.70$ GeV$^2$\,\cite{desy}. These data were fitted
separately by using the CERN-MINUIT code. A combination of
the {\it simplex} and {\it migrad} packages has been used\,\cite{minuit}.
The fitted (modified) cross sections are shown in 
Fig.~\ref{fig:(t-mu2)dsig_dt}, where we can see that, 
except for the DESY data, the $\chi^2/N$ values obtained
from the fits are quite encouraging. 

\begin{figure}[!t]
  \begin{center}
    \leavevmode
    \epsfig{figure=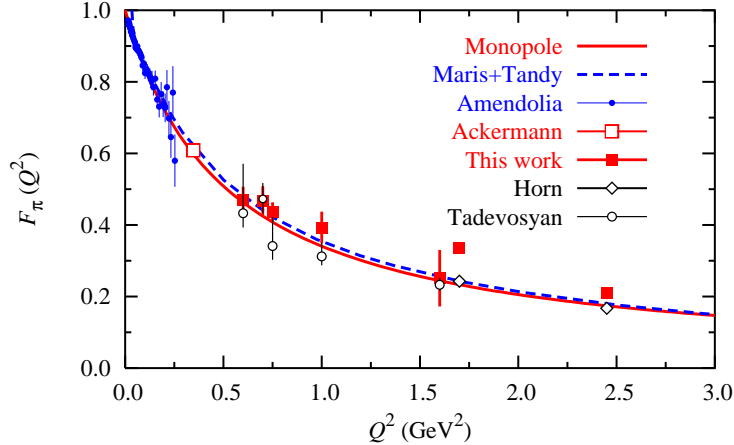,width=100mm}
    \caption{(Color online)  The electromagnetic form factors of pion extracted in this work and
    previous calculations. The monopole form factor is
    obtained by using the pion charge radius of 0.672 fm
    (PDG value~\protect\cite{yao:2006}). 
    The dashed line shows the model of
    Maris and Tandy~\protect\cite{Maris:2000sk}. Results of the present
    work are obtained by using $\Lambda_{\rm h}=0.85$ GeV and shown 
    in this figure by the solid (red) squares.
    Other data shown here are from Amendolia {\it et al.}\,\protect\cite{Amendolia:1986wj}, 
    Ackermann {\it et al.}\,\protect\cite{Ackermann:1977rp}, Horn {\it et al.}\,\protect\cite{Horn:2006tm}, 
    and Tadevosyan {\it et al.}\,\protect\cite{Tadevosyan:2007yd}. The data point
    of Horn {\it et al.} at 1.6 GeV, as well as the corresponding result of 
    the present analysis, 
    have been slightly shifted for visual clarity.
    \protect\label{fig:pion_ff}}
  \end{center}
\end{figure}

The extracted $F_\pi$ for $\Lambda_{\rm h}=0.85$ GeV and 1.30 GeV 
compared with the results of previous 
calculations are listed in Table~\ref{tab:numeical_F_pi}
and depicted in Fig.~\ref{fig:pion_ff}. The 
error bars reported here are also obtained from MINUIT. Their
magnitudes can be easily understood from the comparison between
the fit results (solid and dashed lines) and the experimental data 
as demonstrated in Fig.~\ref{fig:(t-mu2)dsig_dt}. 

\begin{table}
\caption{Values of the extracted $F_\pi$ in the $Q^2$-independent
    analysis obtained by using
    $\Lambda_{\rm h}=0.85$ GeV and 1.30 GeV compared with 
    those of previous works.  
    Reference\,\protect\cite{Tadevosyan:2007yd} has reported ``model
    uncertainties'' listed as the second error-bars in the fourth
    column.\protect\label{tab:numeical_F_pi}}
\begin{center}
\begin{tabular}{cccr}
\hline\hline\\[-2ex]
    ~~$Q^2$~~ & \multicolumn{2}{c}{$F_\pi$ (this work)} & ~~~~$F_\pi$ (previous works) \\
\cline{2-3}\\[-2ex]
    ~(GeV$^2$)~~ &~~ $\Lambda_{\rm h}=0.85$ GeV~~ &~~ $\Lambda_{\rm h}=1.30$ GeV~~ & \\[1ex]
\hline\\[-2ex]
    0.60& $0.470 \pm 0.037$ &  $0.438 \pm 0.066$ &$0.433 \pm 0.017^{+0.137}_{-0.036} $\,\cite{Tadevosyan:2007yd} \\
    0.70& $0.467 \pm 0.041$ &  $0.395 \pm 0.076$ &$0.473 \pm 0.023^{+0.038}_{-0.034} $\,\cite{Tadevosyan:2007yd} \\
    0.75& $0.435 \pm 0.028$ &  $0.396 \pm 0.088$ &$0.341 \pm 0.022^{+0.078}_{-0.031} $\,\cite{Tadevosyan:2007yd} \\
    1.00& $0.393 \pm 0.044$ &  $0.341 \pm 0.074$ &$0.312 \pm 0.016^{+0.035}_{-0.019} $\,\cite{Tadevosyan:2007yd} \\
    1.60& $0.251 \pm 0.079$ &  $0.194 \pm 0.124$ &$0.233 \pm 0.014^{+0.013}_{-0.010} $\,\cite{Tadevosyan:2007yd} \\
    1.60& $0.337 \pm 0.009$ &  $0.276 \pm 0.036$ &$0.243 \pm 0.012 $\,\cite{Horn:2006tm}  \\
    2.45& $0.209 \pm 0.017$ &  $0.153 \pm 0.042$ &$0.167 \pm 0.010 $\,\cite{Horn:2006tm}  \\[1ex]
\hline\hline
\end{tabular}
\end{center}
\end{table}

Figure~\ref{fig:pion_ff} reveals that, within the error bars, 
our simple analysis results in a good agreement with previous
analyses\,\cite{Horn:2006tm,Tadevosyan:2007yd}. At first glance 
the error bars obtained in the present work would seem to be much larger
than those of previous works. However, this is in general not true. For 
example, consider the case of $Q^2=0.6$ GeV$^2$. 
By combining the error bar obtained in Ref.\,\cite{Tadevosyan:2007yd}
with the reported error bar coming from the ``model uncertainty''
we can clearly see that the error bar of the present work
is indeed smaller. For other cases the error bars of the present work
are mostly comparable to those of previous analysis. Only at 
$Q^2=1.60$ GeV$^2$ the obtained error bar is much larger. The
reason behind this behavior is obvious from Fig.~\ref{fig:(t-mu2)dsig_dt}.
Comparing with the result of Ref.\,\cite{Horn:2006tm} at the same
kinematics we can conclude that this error bar can be significantly
reduced if more statistics are available at this point.

The two data points at
1.6 and 2.45 GeV$^2$ from Ref.\,\cite{Horn:2006tm} require a special 
explanation. First, Ref.\,\cite{Horn:2006tm} does not report
the ``model uncertainties'' which appear during the extraction
process. From Table~\ref{tab:numeical_F_pi}
we can see that these uncertainties are substantial in 
increasing the error bars\,\cite{Tadevosyan:2007yd}. 
Second, the discrepancy appears between our result and those
of  Ref.\,\cite{Horn:2006tm} 
for these two data points are presumably due to higher
$W$ and $t$, where other terms excluded in Eq.~(\ref{eq:dsL/dt*(t-m2)})
could start to contribute. Nevertheless, the error bars for 
the two $F_\pi$ at $Q^2=1.6$ GeV$^2$ of our results indicate
that our analysis in this case is still consistent. Therefore, we believe that 
the error-bars obtained in the present work are realistic, especially
if we compare them with those obtained from scattering of 
high energy pions from atomic electrons
at $Q^2\approx 0.25$ GeV$^2$\,\cite{Amendolia:1986wj}. Thus, 
more statistics in the longitudinal cross section data will
reduce these values. 

\begin{figure}[!h]
  \begin{center}
    \leavevmode
    \epsfig{figure=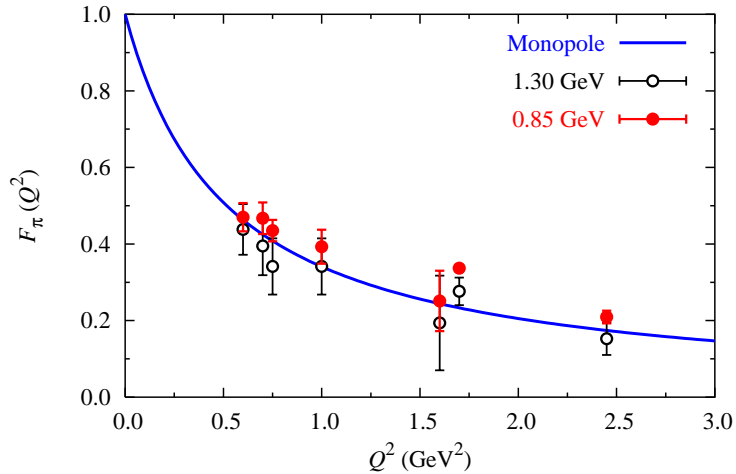,width=100mm}
    \caption{(Color online)  The influence of the variation of the hadronic 
      form factor cut-off $\Lambda_{\rm h}$ on the
      extracted pion electromagnetic form factor.
      The monopole form factor is as in Fig.~\ref{fig:pion_ff}.
   \protect\label{fig:had_var}}
  \end{center}
\end{figure}

In Fig.~\ref{fig:had_var} we demonstrate the variation of the 
extracted electromagnetic form factor $F_\pi$ due to the variation
of the hadronic cut-off ($\Lambda_{\rm h}=0.85$ GeV and 1.30 GeV).
The corresponding numerical values for the two cases are listed 
in Table~\ref{tab:numeical_F_pi}. A soft
cut-off will substantially suppress the cross section. To overcome
this suppression the fit will increase the electromagnetic form factor cut-off.
This explains why the extracted $F_\pi$ values increase once we 
decrease the hadronic cut-off. From this figure we may conclude
that the best agreement with the results of 
Refs.\,\cite{Horn:2006tm,Tadevosyan:2007yd} is obtained 
if we set the 
hadronic cut-off to $1.3$ GeV, a case when we consider
the result of meson exchange models for the nucleon-nucleon
interactions\,\cite{Machleid87}. 

Note that in Figs.~\ref{fig:pion_ff} and \ref{fig:had_var} we
compare our results with the monopole  form factor 
obtained by using the Particle Data Group (PDG)
value\,\protect\cite{yao:2006} of the 
pion charge radius, i.e. 0.672 fm.
In the literature one can find that 
the value ranges from 0.560 fm~\cite{Dally:1977vt}
to 0.740 fm~\cite{Liesenfeld:1999mv}, whereas the recent 
Chiral Perturbation Theory extraction yields a value of
0.661 fm~\cite{Bijnens:1998fm}.

\section{\sc $Q^2$-Dependent Analysis}
\label{$Q^2$_Dependent_Analysis}

In the previous section we have extracted the electromagnetic pion 
form factors at fixed $Q^2$ values. It is however 
possible to analyze the whole experimental data simultaneously 
in a fit. For this purpose, we have to assume that the electromagnetic
form factor can be parametrized with a monopole form factor, 
\begin{equation}
F_{\pi}(Q^2) = \left(1+\frac{Q^2}{\Lambda_{\pi}^2}\right)^{-1}~.
\label{eq:em_ff}
\end{equation}
A quick glance to Fig.~\ref{fig:had_var} indicates that
with this monopole parameterization it would be difficult to produce
the pion charge radius as reported by PDG, especially if we
used $\Lambda_{\rm h}=0.85$ GeV, for which we see that the
two extracted form factors at $Q^2=1.6$ GeV$^2$ and 2.45 GeV$^2$ 
obviously overestimate the monopole parameterization. For 
$\Lambda_{\rm h}=1.30$ GeV we could expect that such
a parameterization would still fairly work, since 
as shown in  Fig.~\ref{fig:had_var} the PDG form factor 
is still within the corresponding error bars. 

Nevertheless,  a simultaneous fit to all available experimental
data is quite important if we want to investigate the internal
consistency of the data or the general trend of the extracted
parameters. Furthermore, within this framework we can leave
the fit to determine not only the electromagnetic form factors $F_\pi$,
but also the hadronic cut-off $\Lambda_{\rm h}$ and 
the pion-nucleon coupling constant 
$g_{\pi NN}$. Here, it is important to note that 
compared to the conventional approaches, which make use
of thousands data points,
a simultaneous fit to only 36 data points is clearly less
reliable to accurately determine the hadronic cut-off and
coupling constant of the pion. Therefore, our motivation 
here is only to confirm the results obtained by
the $Q^2$-independent analysis given in the previous section
as well as to check our claim about the limitation of our
model.

\begin{table}
\caption{Values of some important parameters extracted from fit 
     to three different sets of experimental data in the $Q^2$-dependent
     analysis. References for the literature
     are given in the text.
     \protect\label{tab:fit_par}}
\begin{center}
\begin{tabular}{cccccc}
\hline\hline\\[-2ex]
    Parameter & All data (36 points) & ~~19 Data points~~ & ~~18 Data points~~ & Literature$^\dagger$\\[1ex]
\hline\\[-2ex]
$\langle r_\pi^2 \rangle^{1/2}$ (fm) & $0.576\pm 0.020$ & $0.596\pm 0.111$ & $0.625\pm 0.272$&0.560 - 0.740\\
$g$& $0.019\pm 0.006$ & $0.022\pm 0.013$ & $0.025\pm 0.007$ & - \\
$\Lambda_{\rm h}$ (GeV) & $0.850\pm 0.034$ &$0.968\pm 0.256$ & $1.300\pm 0.230$ & 0.400 - 1.500\\
$g_{\pi NN}^2/(4\pi)$ & $14.500\pm 0.827$~ & $14.500\pm 0.623$~ & $14.500\pm 0.721$~ & 13.45 - 14.52\\
$\chi^2/N$ &1.118 &0.348 &0.263 & - \\[1ex]
\hline\hline\\[-2ex]
$^\dagger$See text.
\end{tabular}
\end{center}
\end{table}

\begin{figure}[!t]
  \begin{center}
    \leavevmode
    \epsfig{figure=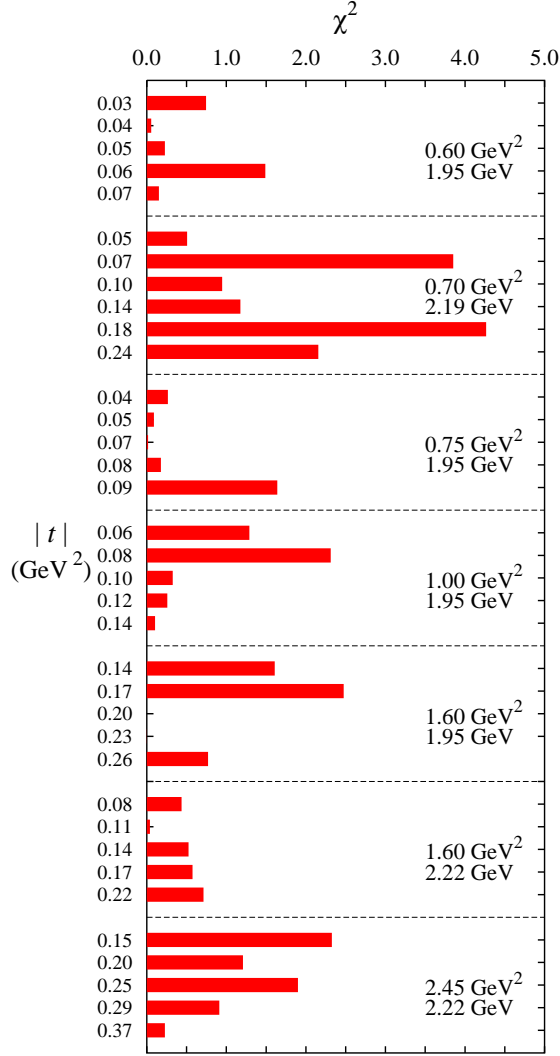,angle=-90,width=75mm}
    \caption{(Color online)  Values of the $\chi^2$ obtained from fit
         to all experimental data in the $Q^2$-dependent analysis
         plotted as a function of $|t|$.
         Note that the two numbers given in each panel show the
         corresponding values of $Q^2$ and $W$, respectively. 
         The order of the experimental data sets shown in this
         figure (from top to bottom) is as in Fig.~\ref{fig:(t-mu2)dsig_dt}.
   \protect\label{fig:chi1}}
  \end{center}
\end{figure}

As a first step, we fit all experimental data by leaving
the pion charge radius $\langle r_\pi^2 \rangle^{1/2}$ 
(for the sake of comparison,
instead of the $\Lambda_\pi$), the
constant $g$ in Eq.~(\ref{eq:fit_function}), the hadronic
cut-off $\Lambda_{\rm h}$, and the coupling constant $g_{\pi NN}$,
as free parameters. Since we only intent to qualitatively study
the extracted parameters, we limit the pion charge radius in
the range of 0.560 fm - 0.740 fm, the hadronic cut-off between
0.85 GeV and 1.30 GeV, and the pion-nucleon coupling constant
$g_{\pi NN}^2/(4\pi)$ between 13.50 and 14.50.
The corresponding numerical values extracted from
this fit along with their $\chi^2/N$ are shown in the second column 
of Table~\ref{tab:fit_par}. The individual $\chi^2$ as a function of 
$t$ is plotted in Fig.~\ref{fig:chi1}. Fitting to all available data
results in a total $\chi^2$ of about 36.
Interestingly, we found that more than 36\% of it come from
the reanalyzed DESY data at $W=2.19$ GeV, $Q^2=0.70$ 
GeV$^2$\,\cite{desy}. This indicates that these data are probably 
incompatible with the recent JLab data \cite{Horn:2006tm,Tadevosyan:2007yd}.
Furthermore, we found that these data belong also to the high-$t$ and $W$ data.

The more surprising result is however shown by the set of the new JLab data 
at $Q^2=2.45$ 
GeV$^2$\,\cite{Horn:2006tm}. Although this data set contains
five precise data points, the combination of high $t$, $W$,
and $Q^2$, obviously puts these data beyond the limitation of our approach.
For $Q^2=0.60$ GeV$^2$ only the data point
at $|t|=0.06$ GeV$^2$ yields $\chi^2> 1$. This is understandable
if we look at the first panel of Fig.~\ref{fig:(t-mu2)dsig_dt},
where it is obvious that this data point slightly underestimates
the trend of the other data. The same situation happens for
the last data point at $Q^2=0.75$ GeV$^2$.
Finally, we found that 
the first two data points at $Q^2=1.00$ GeV$^2$ and
$Q^2=1.60$ GeV$^2$ overestimate the fit result. This fact
is relatively difficult to explain, especially for the
first case, for which the corresponding $t$ are clearly
small. Probably, these data points are too high or other
physics beyond our approach at this kinematics are missing.

As a second step we remove the data points that produce
$\chi^2 > 1$ from our fitting data base. The number of
data points at this step turns out to be 19. The extracted parameters
obtained from fitting to these data are shown in the third 
column of Table~\ref{tab:fit_par}. Obviously the $\chi^2/N$
decreases significantly. The pion charge radius and hadronic
cut-off increase slightly from those of the first fit, whereas
the pion coupling stays the same. The only interesting result at this
step is the fact that the highest $\chi^2$ comes from the data
point at  $W=2.22$ GeV, $Q^2=1.60$, $|t|=0.22$ GeV$^2$. Again,
this is understandable because of the combination of high $t$
and $W$ and, furthermore, this data point seems to underestimate
the trend of other four data points. 

\begin{figure}[!t]
  \begin{center}
    \leavevmode
    \epsfig{figure=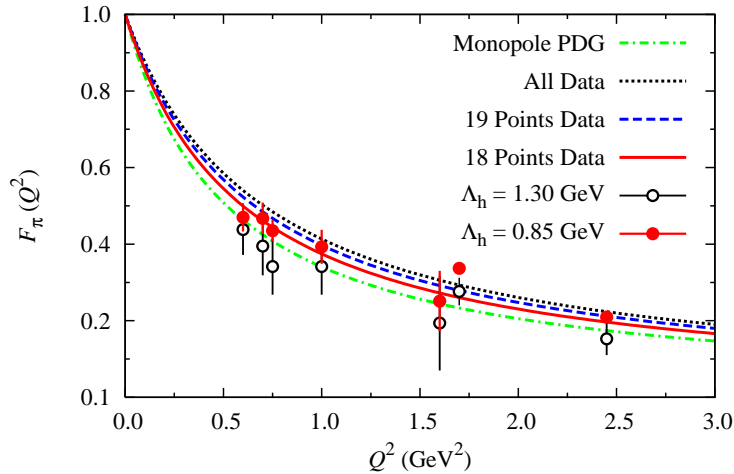,width=100mm}
    \caption{(Color online) Electromagnetic form factors of the pion
         according to the results of fitting to three different
         sets of experimental data in the $Q^2$-dependent analysis. 
         The corresponding pion charge
         radii are listed in Table~\ref{tab:fit_par}. For comparison, 
         results of
         the $Q^2$-independent analysis with $\Lambda_{\rm h}=0.85$
         GeV and 1.30 GeV are also shown by the solid and open
         circles, respectively.
   \protect\label{fig:monopole_fit}}
  \end{center}
\end{figure}

As the last step we remove the data point at   $W=2.22$ GeV, $Q^2=1.60$, 
$|t|=0.22$ GeV$^2$ from our fitting data base and refit the rest of the 
data (18 points).
The extracted parameters for this step are shown in the fourth 
column of  Table~\ref{tab:fit_par}. The value of $\chi^2/N$ is
very small, indicating that our approach works pretty well for
these selected data. The pion charge
radius (as well as its error bar) and the hadronic cut-off
increase significantly after we have removed the last data
point as described above. Although having a relatively large error
bar, the pion charge radius is now much closer to the PDG value
compared to those obtained in the first and second steps. Thus,
the result of the third step corroborates the finding in 
the $Q^2$-independent analysis, i.e., smaller $\chi^2$ values would
be obtained if we used $\Lambda_{\rm h}=1.30$ GeV. Another
way to understand this result is by comparing the electromagnetic
form factors $F_\pi$ obtained from this analysis with that of
the PDG and those obtained from the $Q^2$-independent analysis
described in the previous section. This is shown in 
Fig.~\ref{fig:monopole_fit}, where we can clearly see that
omitting the high-$\chi^2$ data (put in other words, using 
experimental data with the kinematics allowed by the model)
results in a form factor that is in closer agreement with
the form factor of PDG and those obtained from the 
$Q^2$-independent analysis with $\Lambda_{\rm h}=1.30$ GeV.

It is also important to note that the values of $g$
and pion coupling constant are relatively stable
for the three different fits
(see Table~\ref{tab:fit_par}). The relatively large
error bars for the latter indicate that our 
conclusion drawn from the two analyses 
is not too sensitive to the variation 
of the $g_{\pi NN}$ within the fit limit.

\section{\sc Conclusion}
\label{Conclusion}
In conclusion, we would like to say that we are able 
to extract the electromagnetic form factor of 
pion from the recent longitudinal cross section data of 
pion electroproduction
from JLab by using a simple $t$-channel diagram, which utilizes
fewer assumptions compared to the previous calculations. 
For this purpose we have performed a $Q^2$-independent analysis 
supplemented with a $Q^2$-dependent analysis.
The result of the first analysis shows a good 
agreement with $F_\pi$ extracted from previous 
works. The extracted values depend on the choice of the hadronic
form factor cut-off. Nevertheless, by using the generally accepted value we 
still obtain a satisfactory result. The best result would be obtained if 
we used a cut-off from the meson exchange model. 
The result of the second analysis corroborates
the findings of the first one. 
More accurate, with 
also smaller $|t|$, longitudinal cross section data will certainly help 
to reduce the uncertainty of the present calculation. Finally, we would
like to mention that this simple calculation provides for the first
time a direct proof that at the given kinematics pion electroproduction is
dominated by the $t$-channel process. Previous works considered this
merely as an assumption. This result could set a new constraint on the
phenomenological models that try to explain the process, i.e. at the
kinematics given in this paper contributions from other channels
should vanish or a least should be minimal. We note that, however, 
at small $t$ but
higher energies, contributions of certain resonances to the process
could be substantial.

\section*{\sc Acknowledgment}
The author acknowledges the support from the University of
Indonesia.

\end{document}